# Noise induced new quasi-period and periods switching

Yuxuan Wu[1*], Yuxing Jiao[1†], Yanzhen Zhao[2‡], Haojun Jia[3§] and Liufang Xu[1**]

[1]*Department of Physics and Biophysics & Complex System Center,
Jilin University Changchun 130012, P. R. China*

[2]*Department of Physics, Applied Physics and Astronomy, Rensselaer Polytechnic Institute,
Troy, New York 12180, USA*

[3]*Department of Chemistry, Massachusetts Institute of Technology, Cambridge, MA 02139*

We employ a typical genetic circuit model to explore how noise can influence the dynamic structure. With the increase of a key interactive parameter, the model will deterministically go through two bifurcations and three dynamic structure regions. We find that a new quasi-periodic component, which is not allowed by deterministic dynamics, will be generated by noise inducing in the first two regions, and this quasi-period will be more and more stable along with the noise increasing. Especially, in the second region, the quasi-period will compete with a stable limit cycle and perform a new transient rhythm. Furthermore, we ascertain the entropy production rate ($EPR$) and the heat dissipation rate ($HDR$), and discover a minimal value with elucidating theoretically. In the end, we unveil the mechanism of the quasi-periods forming, and show a practical biological instance. We expect that this work is helpful to solve some biological or ecological problems, such as genetic origin of periodical cicadas and population dynamics with fluctuation.

## Introduction

Periodicity of life has always been a fascinating issue. This periodicity is often realized by a set of genetic circuits, which is ubiquitous in life. Many important biological functions, such as circadian rhythms of mammals and some unicellular organisms [1-3], cell cycle processes of bacteria [4], oscillators in Xenopus embryonic cells [5], etc., are regulated by a genetic circuit containing positive and negative feedback. The essence of the genetic circuit is some interrelated biochemical reactions, which can be described by a set of differential equations.

However, the biosystem cannot be simulated only by deterministic dynamics, because noise is inevitable, even a natural and indispensable part of biological systems, having brought a lot of interesting phenomena. For instance, the regulation of noise effect in eukaryotic gene expression [6], stochastic focusing phenomenon [7], the influence of noise in Bacillus subtilis cells [8], the positive roles in plant cells [9], and many other discoveries [5,10-15]. Noise can also have the more counterintuitive effect of generating new stable states that do not exist in the absence of fluctuations, and most of the existing works focus on the two-state transition induced by noise [16-20]. This transition requires the bistable structure of the system. Further, It is of practical significance to explore whether random action can produce near stable periodic oscillation, to which have been some theoretical efforts on this matter [21,22]. If so, what kind of new system will be formed by the new periodic behavior and the original system? What will happen to its nature? Hitherto, there is no definitive answer to these questions.

Our work takes a concrete model as the starting point and endeavors to answer the above questions, of which the model was first proposed by literature [23]. It is a typical genetic circuit structure with positive and negative feedback, which often appears in various biological systems (in some cases as a part of the system) [4,24]. We elaborate it in the chapter for introducing the model, and show its detailed dynamic properties, as well as an undiscovered Hopf bifurcation point, which leads to the emergence of a stable limit cycle state. By changing a key parameter, the system will experience three different dynamic structures. A new quasi-periodic dynamic state induced by noise appears in two of the three dynamic structures. Particularly, this new periodic structure is impossible to exist according to deterministic dynamics.

With the help of the theory of generalized potential and probability flow [25,26], we show the dynamic structure of the system in the form of the probability distribution, and the motion trend of the state points representing the position of the system in the phase space. The mutual corroboration of the above two can be confirmed by the analysis of the trajectory of the state point in the noise environment. Next, we define the barrier height and the average first pass time, the former can quantify the ability of the system to form a new periodicity, and the latter can effectively measure the cycle length of this periodicity. We have also clarified the relationship between the two. Especially, we give a practical biological example and make a brief analysis to show the possible universality of this noise-induced new periodicity in biology.

---
[*] yxwu19@mails.jlu.edu.cn
[†] jiaoyx1118@mails.jlu.edu.cn
[‡] zhaoy22@rpi.edu
[§] haojun@mit.edu
[**] lfxuphy@mails.jlu.edu.cn

Table.1. Properties of system shown in FIG.1.in three parts of $k_2 \in [0.12, 0.17]$

| $k_2$ | [0.12, 0.1474] | [0.1475, 0.1559) | (0.1559, 0.17] |
|---|---|---|---|
| Intersection (a) | Stable fixed point | Stable fixed point | Stable fixed point |
| Intersection (b) | Saddle | Saddle | Saddle |
| Intersection (c) | Unstable focus | Unstable focus | Stable focus |
| Notes | | Stable limit cycle around the focus | |

Additionally, the global properties of the system are also what we want to grasp. In general, due to the non-equilibrium nature of biological systems, entropy generation rate is a more common physical quantity in related research [27-29]. Indeed, the determination of entropy production rate plays a decisive role in active particles [30,31], single-cell biological systems [32], soft biological materials [33], complex networks [34], and so on. Equally, the concomitant heat dissipation rate is also an important physical quantity [35-39]. When describing the global properties of the non-equilibrium system, they are like two sides of a coin. Some studies also show that the entropy production rate and heat dissipation rate are intimately related to the robustness of the system [40-42].

We offer an interpretation for the physical significance of the minimal value we found in the research of the heat dissipation rate. One can see *SI text* for a proof for consolidating the physical significance we interpret, and an interesting intuitive result.

## Model

### Picture determined by nonlinear dynamics and the changes in topology

The model can be simplistically described by a set of two-dimensional equations as

$$\frac{dx}{dt} = F_1(x, y) \qquad (1)$$

$$\frac{dy}{dt} = F_2(x, y) \qquad (2)$$

in which $F_1 = \alpha_1 + \beta_1 x^n/(k_1^n + x^n) - \delta xy - \lambda_1 x$ and $F_2 = \alpha_2 + \beta_2 x^p/(k_2^p + x^p) - \lambda_2 y$. This set of equations describe the dynamics behavior of a biological system in phase space. One can see details of coefficients in $F_1$ and $F_2$ in Appendix. In this model, the change of dynamic properties and topological structure is obtained by changing $k_2$. We focus on the kinetic behavior in the first quadrant. In the range $k_2 \in [0.12, 0.17]$, nullclines of $x$ and $y$ have three intersection points, and two bifurcation points of the dynamics structure (K1 and K2) were found, as what were shown in Fig.1 and Table.1. The reason why we choose the range $k_2 \in [0.12, 0.17]$ is that it is symmetric to the two bifurcation points which have a critical influence on the structure of the system, and the system will maintain the corresponding structure outside this range. The stability of those three intersections can be determined by discriminants. One of those two bifurcation points (K2), $k_2 = 0.1559$, can be determined by solving nullclines ($F_1 = 0$ and $F_2 = 0$) with discriminant of focus simultaneously. Another one (K1), $k_2 \in (0.1474, 0.1475)$, was found by simulation but not analysis. What should be emphasized is that we found a stable limit cycle around the unstable focus in range $k_2 \in [0.1475, 0.1559)$.

Bifurcation of the system means the changes of topological structure [43]. In nonlinear dynamics, we consider it as the phase transition of the system. Stable and unstable manifolds are streamlines starting from or going back to saddle. Therefore, we can judge and predict the behavioral trend of every streamline by considering the shape of nullclines together with manifolds. Markedly, there is a bifurcation (K1) on the number axis between $k_2 = 0.1474$ and $k_2 = 0.1475$, on each side of which the structure of the system is essentially different from the structure on another side (Fig.S1 (d,e,d1,e1) in *SI text*). Fig.S1(f,f1) shows a Hopf bifurcation at $k_2 = 0.1559$ (K2). From the right to the left of K2 on the number axis, a stable focus develops into the stable limit cycle, which is the typical characteristic of Hopf bifurcation.

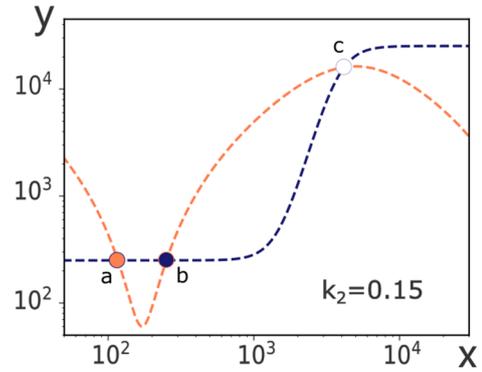

FIG.1. Nullclines of $x$ and $y$, and their intersection, choose $k_2 \in [0.12, 0.1474]$ as a sample. The orange dash line depicts x nullcline, and blue for y. Intersection depicted by the solid circle filled with orange(a) is a stable fixed point, solid circle filled with blue(b) for the saddle, and open circle(c) for focus. In the range $k_2 \in [0.12, 0.17]$, the system will maintain this structure.

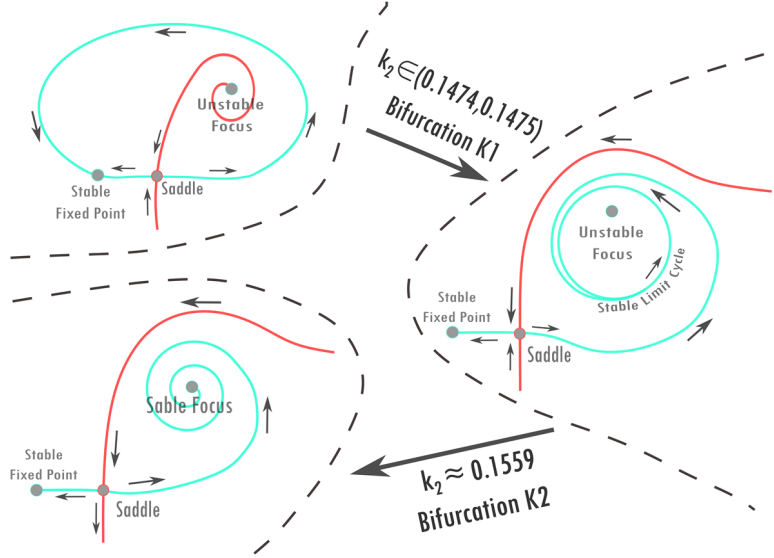

FIG.2. Summary illustration of the topological structure's change. Crimson solid lines depict the stable manifolds And cyan solid lines for unstable manifolds, and trends of manifolds are marked by arrows. As $k_2$ crosses K1, the stable manifold is peeled off from the focus, and a node that started at the stable fixed point strides over the part between the stable fixed point and the saddle. Meanwhile, there is a stable limit cycle formed. After $k_2$ crosses K2, the limit cycle shrinks to the focus. It's like a donut turning into a mark cup.

Sum up the global kinetic properties determined by deterministic nonlinear dynamics in the first quadrant (more details about manifold and streamline of $k_2$ being selected as different values in *SI text*):

① $k_2 \in [0.12, 0.1474)$. The system has a fixed point in the lower-left part nearby the saddle, which is the only stationary structure of the system. Unstable manifold forms a large closed loop, by which the unstable focus is encircled. That hints our system can form a loop if the state point of the system at the lower-left fixed point strides across the saddle.

② $k_2 \in (0.1559, 0.17]$. The whole region is divided into two parts by a stable manifold back to the saddle, in each of which there is a stationary structure being the stable focus and the stable fixed point. In the area where the stable fixed point is located, every state point will return to the stable fixed point. Similarly, state points will return to the stable focus in another area.

③ $k_2 \in (0.1475, 0.1559)$. The whole region is divided into two parts by a stable manifold similar to what we see in ②, in each of which there is a stationary structure being the limit cycle and the fixed point. If $k_2$ is close to the right neighborhood of K1, the limit cycle will cling to the stable manifold being a large loop, which is similar to that in ①. If $k_2$ is close to the left neighborhood of K2, the limit cycle will tightly surround the unstable focus in the neighborhood of the focus, which is similar to that in ②.

### Stochastic Dynamics Model

We construct a stochastic dynamic model by adding white noise to the deterministic nonlinear dynamic system, for the reason that white noise is a good approximation to the actual situation, and this simple noise is very representative [44]. The stochastic differential equations (SDE) derived from eq. (2) and (3) is written as:

$$\frac{dx}{dt} = F_1(x,y) + \sqrt{2D} \cdot \eta(t) \qquad (3)$$

$$\frac{dy}{dt} = F_2(x,y) + \sqrt{2D} \cdot \eta(t) \qquad (4)$$

in which the definitions of $F_1$ and $F_2$ are the same as what are in the previous section. From the point of view of dynamics, this is a set of overdamped Langevin equations. $D$ is the constant and isotropic diffusion coefficient, meaning the off-diagonal elements of diffusion matrix $\boldsymbol{D}$, $D_{12}$ and $D_{21}$, is zero. $\eta(t)$ is the standard Gaussian white noise, $\langle \eta(t)\eta(t') \rangle = \delta(t-t')$, which can be regarded as a Wiener process with a diffusion coefficient of 1. According to these definitions, the global kinetic property can be described by a Fokker-Planck (F-P) equation derived from eq. (3) and (4) [44]:

$$\frac{\partial P(x,y,t)}{\partial t} = -\frac{\partial}{\partial x}(F_1 P) - \frac{\partial}{\partial y}(F_2 P) + D\left(\frac{\partial^2}{\partial x^2} + \frac{\partial^2}{\partial y^2}\right)P \qquad (5)$$

The F-P equation shows how our nonequilibrium system's state points distribute in phase space by probability. But what we are more interested in is the motion of the system in steady state. Eq. (5) can be transformed into

$$\frac{\partial P(x,y,t)}{\partial t} = -\nabla \cdot \boldsymbol{J} \qquad (6)$$

in which $J_1 = F_1 P - D(\partial P/\partial x)$ and $J_2 = F_2 P - D(\partial P/\partial y)$. At the long time limit, our nonequilibrium system reaches steady state, $\partial P/\partial t = -\nabla \cdot \boldsymbol{J} = 0$. The divergent free steady state probabilistic flux is rotational, which means $\boldsymbol{J}$ does not have to vanish and can be expressed as

$$\boldsymbol{J}_{ss} = \boldsymbol{F} P_{ss} - \boldsymbol{D} \cdot \nabla P_{ss} \qquad (7)$$

As the steady state probability flux vector, $\boldsymbol{J}_{ss}$ can represent the state points' motion trend of the system in the non-equilibrium steady state (NESS). Therefore, dividing both sides by $P_{ss}$, one gets

$$\boldsymbol{F}' = \frac{\boldsymbol{J}_{ss}}{P_{ss}} = \boldsymbol{F} + \boldsymbol{D} \cdot \nabla U \qquad (8)$$

in which $U = -\ln(P_{ss})$ as generalized potential, and $\boldsymbol{F}'$ is composed by the deterministic generalized driving force $\boldsymbol{F}$ and diffusion effect [45]. In our work shown in the later section, we use $U$ as the potential landscape to reveal the structural changes of the system in NESS.

The deterministic dynamic mechanism of the system ensures that the trajectory from a certain starting point will concentrate in the range of a certain path without excessive deviation when the noise is not too large. That 'certain path', because of the presence of noise in stochastic dynamics circumstances, will be a little different from what in deterministic dynamics. We used stochastic path integral to determine a dominant path, which benefitted our work [46]:

$$\underset{for\ all\ paths\ \boldsymbol{x}(t)}{\int \cdots \int} P[\boldsymbol{q}(t)] \mathscr{D}\boldsymbol{q}(\boldsymbol{t}) \qquad (9)$$

in which $\mathscr{D}\boldsymbol{q}(t)$ is integral for all paths and $P[\boldsymbol{q}(t)]$ is the probability of path $\boldsymbol{q}(t)$ in phase space:

$$P[\boldsymbol{q}(t)] = \exp\left(-\int_{t_a}^{t_b} \mathcal{L}[q(t), \dot{q}(t)] dt\right)$$

$$= \exp\left\{-\int_{t_a}^{t_b} dt \left(\frac{1}{4D}[\dot{q}(t) - f(q(t))]^2 + \frac{1}{2}\frac{df(q(t))}{dq(t)}\right)\right\} \qquad (10)$$

The $f(q(t))$ is the deterministic driving force and $\mathcal{L}[q(t), \dot{q}(t)]$ is the lagrangian of the system. Using the Euler-Lagrange equation, the dominant path can be derived [47].

Besides, to ensure consistency, we mainly adopt the direct discrete simulation method to solve the SDE. Nevertheless, for a model based on a practical genetic circuit realizing its function by biochemical reaction, we also run the Gillespie stochastic simulation algorithm to support our work. Details in *SI* text.

## Result and discussion

### Landscape and streamlines plotting

From a micro point of view, phenomena, which appear in a system determined by equations with the form like eq.(3) and (4), are the joint presentation of cooperation and competition between stochastic noise and deterministic generalized forces. In our work, the dynamic structure in phase space is reconstructed by noise, or so-called fluctuating force, resulting in a new physical state of the global system in phase space throughout the entire domain of $k_2$. In order to further study this new property, we select a representative point in each of the three domains of $k_2$. (Fig.3) is the deterministic dynamics structure of them.

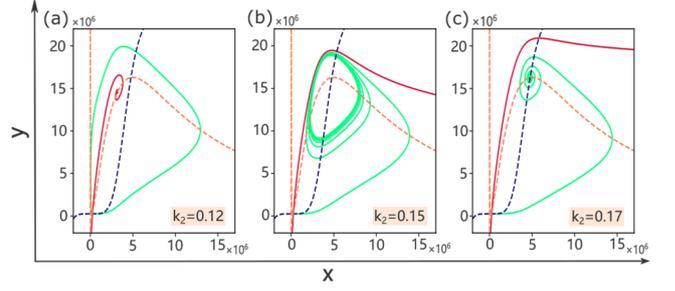

FIG.3. Nullclines, stable and unstable manifolds of representative points in each of the three domains of $k_2$. Springgreen solid lines are unstable manifolds, and magenta solid lines are stable manifolds. Indigo dash lines are nullclines of y, and orange dash lines are nullclines of x.

As we have defined in the previous section, landscapes and steady state probability fluxes on it depict the kinetic structure of the system. Noticeably, a new large circle adhered to the unstable manifold emerges, which can never exist in the deterministic dynamics system (Fig.4). It is not only because this phenomenon is completely the product of stochastic action, but more importantly, the large circle exists according to the Bendixson-Dulac theorem if it is regarded as a limit cycle to some extent.

The streamlines plotting can be used as the evidence of the conclusion of landscapes (Fig.5). For the situation $k_2 = 0.12$, the large circle including of the stable fixed point and the saddle constitutes all of the dynamics structure. It is because that all of the state points in phase space will return to the stable fixed point driven by the deterministic generalized force, and the white noise makes these state points cross the saddle with a certain probability. For the situation $k_2 = 0.15$, this complex system has two circles. One is the stable limit cycle we already verified, and another is the large circle created by the noise. Obviously, a part of the limit cycle clings to the stable manifold returning to the saddle point (Fig.3(b)). Because of this, the state points running on the limit cycle can easily cross the stable manifold driven by noise (This is also one of the reasons why $k_2 = 0.15$ is chosen as the representative value). This means that the two circles may switch alternately. Besides, the larger the diffusion coefficient $D$ is, the more the two circles mingle with each other, which is because that the state points already jumped into the large circle have a higher probability of jumping back to the right side of the stable manifold before returning to the stable fixed point. As for $k_2 = 0.17$, the stable manifold is far away from the unstable manifold. Compared with the former two, it is relatively difficult to form a large circle in this case. Therefore, we do not pay much attention to this situation, as long as we know that there are still transitions between two stable structures in the form of the large circle.

With the increase of the diffusion coefficient $D$, for $k_2 = 0.12$, the appearance of the large ring will change from occasional pulse behavior to quasi-periodic behavior (Fig.6). In the biological system we study, it implies an oscillator whose period is distributed in a certain range. For $k_2=0.15$ with the increase of $D$, the amplitude of the x-axis, which is of smaller oscillation representing the limit cycle, gradually becomes unstable and appears on a lower and lower frequency. The frequency of the appearance of the amplitude which represents the large circle, being contrary to the former, gradually increases. The frequency of both parties gradually reached a relatively fixed ratio (Fig.7). This phenomenon can be regarded as the continuous phase transition of the system with the change of diffusion coefficient $D$.

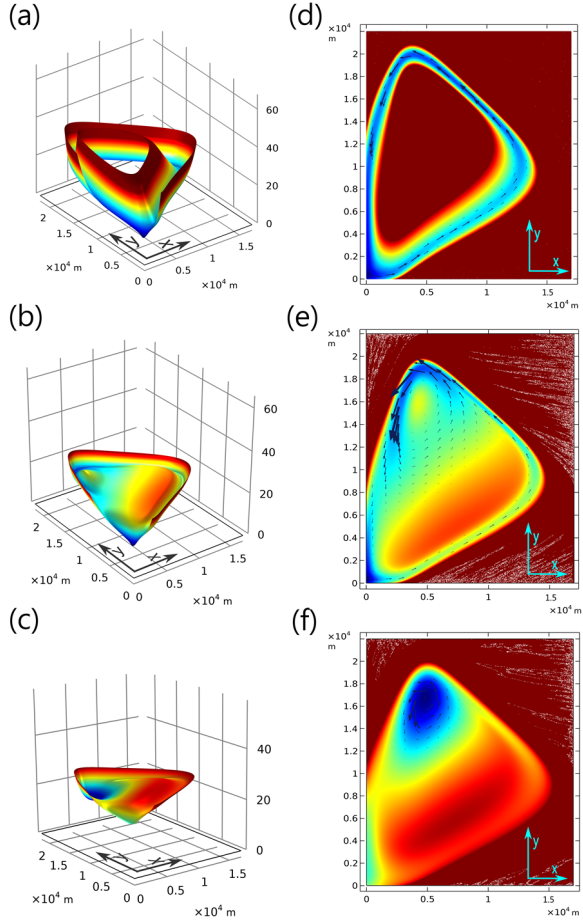

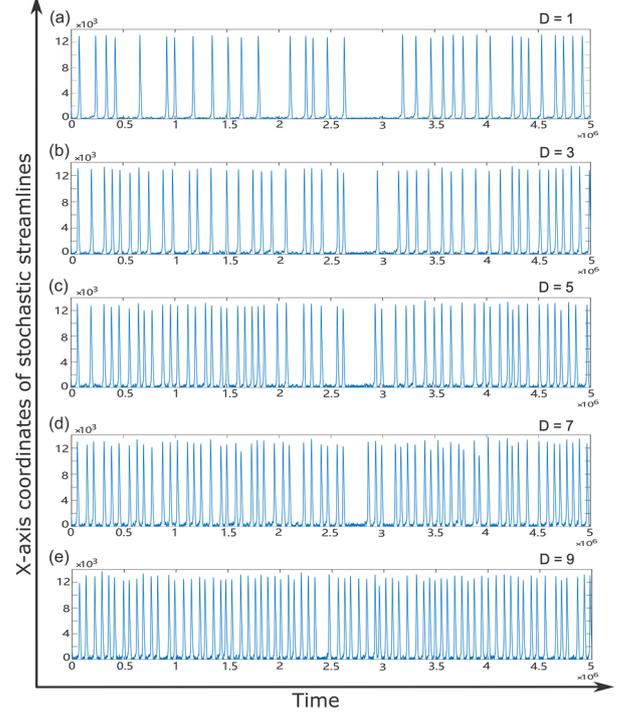

FIG.4. Landscape and vertical view of $k_2=0.12, 0.15$ and $0.17$. (a-c) are landscapes of $k_2=0.12, 0.15$ and $0.17$ with $D=5, 5$ and $30$. The pruned parts are areas where the probabilities approach zero. (d-f) are vertical views of (a-c), in which ranges of color change are the same with (a-d). The dark blue arrows in vertical views show the $J_{ss}$, the length being proportional to the sizes of the flow at points they are drawn. These arrows may not express the values of fluxes precisely, but they can be used to characterize where there are significant probability flows. More figures depicting the influence to the dynamic structures by changing $D$ are in *SI text*.

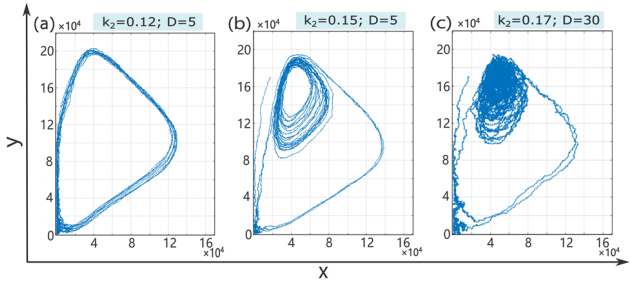

FIG.5. Stochastic streamlines plotting of $k_2=0.12, 0.15$ and $0.17$. The starting points of these streamlines are all $(1750, 17000)$.

## Barrier height, first pass time and competing capacity of the new quasi-period

From Fig.7 and 8, we can see that the widths of the thin and high peaks representing the large circle have no obvious change. Therefore, almost all the differences, which are between periods of large circles with different

FIG.6. Evolution of x-axis coordinates with time for $k_2=0.12$. horizontal ordinates of (a-e) are $x$, and longitudinal coordinates of (a-e) are $y$. Since the dynamics of the system takes the form of oscillators, these diagrams look like amplitude diagrams. These graphs can be regarded as a supplement to the stochastic streamline plotting, which is convenient for us to observe the quasi-periodic behavior of the system

diffusion coefficients $D$, come from the expected time required to cross the barrier at the saddle from the vicinity of the stable fixed point. Furthermore, only crossing the saddle does not guarantee that the state point will continue to move along with the unstable manifold and form the large circle. This can be learned from the streamlines plotting and t-x plotting, because the state point still has the possibility of returning to the left side of the saddle after crossing the saddle. In the vicinity of the line between the saddle and the stable fixed point in the lower-left corner, the stochastic force is more dominant than the deterministic generalized force. To sum up, the mean first passage time (MFPT) from the stable fixed point to a certain point on the right side of the saddle can be used as a scale to measure the period of the large circle, and to some extent, it can also be regarded as the expected value of the period of the large circle (Fig.8(a, b)). In our work, the mean first passage time is not calculated by formula, but by directly averaging the first passage time obtained from 10000 measurements. The definition of a specific exit point (FXP) is explained below.

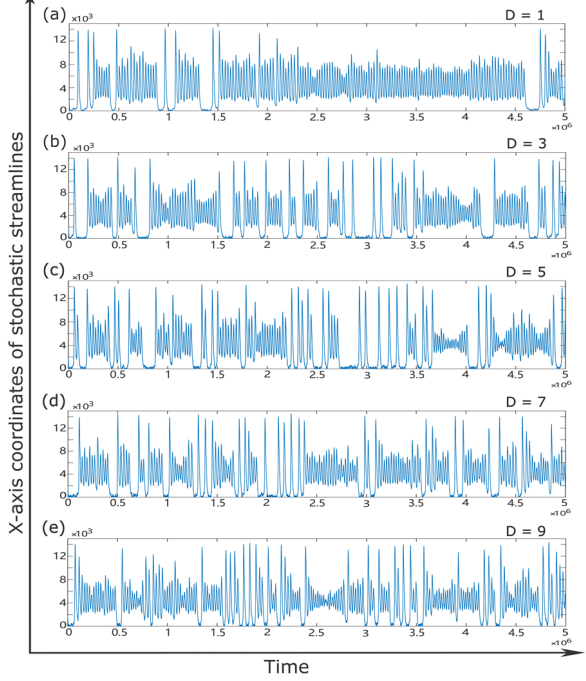

FIG.7. Evolution of x-axis coordinates with time for $k_2 = 0.15$. horizontal ordinates of (a-e) are $x$, and longitudinal coordinates of (a-e) are $y$

In Fig.8(a, b), we have seen that with the increase of the diffusion coefficient $D$, the mean first pass time decays rapidly and then tends to be stable and flat. This corresponds to Fig.6 and Fig.7, that is, the distance between peaks becomes closer with the increase of diffusion coefficient $D$. Then, another problem is how to ensure that the state point under the effect of noise will cross the saddle and move along the unstable manifold with a certain probability, instead of diffusing to other areas. As we mentioned earlier, calculating the dominant path by (eq.10) is a line and efficient solution, which is also the basis for defining the barrier height hereinafter.

As we see in Fig.9, In the region where we are concerned and define the first passage time and barrier height, the main path almost coincides with the unstable manifold. When the total time steps are the same, the probability of the dominant path is much greater than that of other paths. This fact can also be seen by comparing (Fig.5) with (Fig.9). Even if there is noise, the streamlines will not deviate from the dominant paths.

Supported by these conclusions, we define the barrier height as follows: Barrier Height $\triangle H = U_{FXP} - U_{Fixed}$. As the definition, the value of barrier height is the difference between the value of landscape $U$ at FXP and that at the stable fixed point. The definition of FXP ensures the unity between barrier height and first passage time, which is convenient for us to investigate the relationship between them. The data is shown in Fig.8(c, d).

Overviewing Fig.8(a-d), we can draw the following conclusions. First, the MFPT and the barrier height decrease monotonically with the increase of diffusion

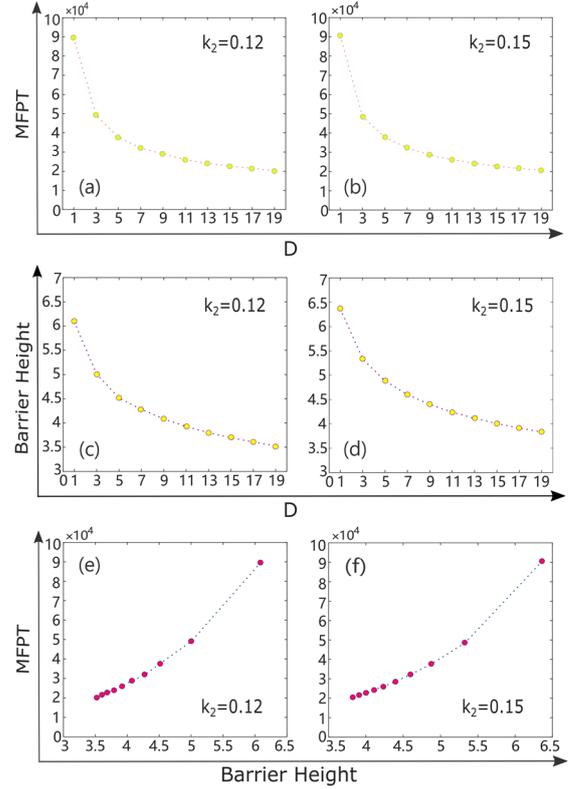

FIG.8. (a, b) Mean first pass time of $k_2 = 0.12$ and $k_2 = 0.15$. The state point moves to start from the stable fixed point in each measurement, and we confirm one measurement has been accomplished if the x-axis coordinate of the state point reaches the x-axis coordinate of the unstable focus. For $k_2 = 0.12$, the coordinate of the unstable focus is (3179, 14547), and for $k_2 = 0.15$, the coordinate of the unstable focus is (4171, 15998). One of the reasons for setting the boundary conditions is that when the state point reaches the x-axis coordinate of the focus, the possibility of returning to the saddle will become extremely small so that such a situation is unlikely to be observed in a limited time. Another reason is that it can be compared with the analysis of barrier height in the following paper. Moreover, in (b), $k_2 = 0.15$, we reduce the interval between the points of diffusion coefficients $D$ from 1 to 10, for the reason that it can be compared with the analysis of heat dissipation in the following. (c, d) Barrier height of $k_2 = 0.12$ and $k_2 = 0.15$. (e, f) The relation between barrier height and MFPT.

coefficient $D$. It is not difficult to understand that with the increase of diffusion coefficient, the 'activity' of the state point increases. In other words, the width of the expected value of the white noise composed by the Wiener process increases, which can cause appreciable effects. This is reflected in the landscapes, where the global shape flattens (Fig.4). Second, with the increase of the diffusion coefficient $D$, the mean first pass time decays rapidly and then tends to be stable and flat. Observing the solution of a general Wiener process may help us to understand this fact that only when the diffusion coefficient $D$ is small, the shape of the function of the solution will change significantly with the change of diffusion coefficient $D$. Third, The polyline of barrier heights and the MFPT have similar shapes. In fact, there is also a monotonic relationship between the MFPT and the barrier height, in which the barrier height acts as a scale to measure the diffusion coefficient $D$ (Fig.8(e, f)). This result is similar to that in reference [26]

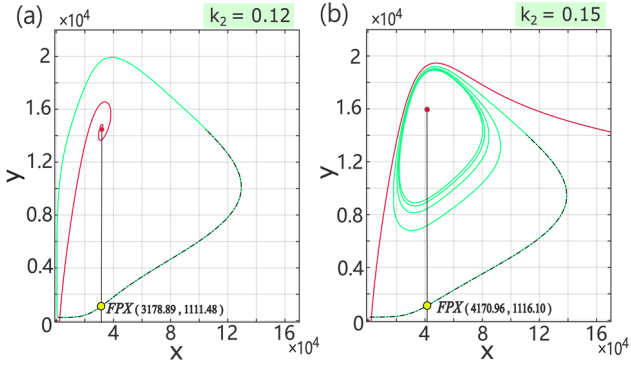
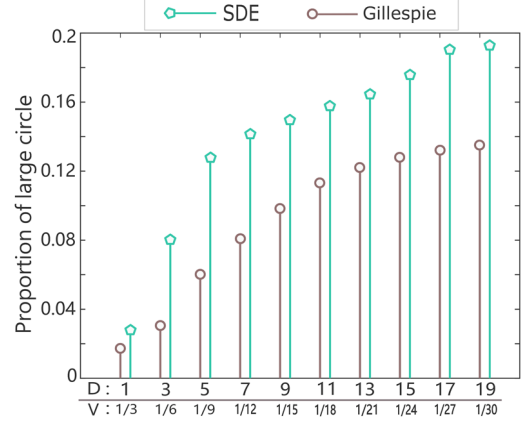

FIG.9. Manifolds plotting and dominant paths. We select the situation $D=5$ as examples. The use of colors and line types of manifolds are the same as (Fig.3). The black dash lines are the dominant paths calculated by (eq.10). Because the dominant path integral will diverge in the region where the unstable manifold and the stable manifold are close to each other (This fact can also be seen in figures of landscapes that the landscapes tend to be flat in this area (Fig.4)), we truncate the total step length of the dominant path integral. Each dominant path starts from the stable fixed point. For $k_2=0.12$, the stable fixed point locates at (116.611, 250.002). For $k_2=0.15$, it is (116.611, 250.001). FPX is the intersection of the vertical line made by the focus to the x-axis and the dominant path. For $k_2=0.12$, the unstable focus locates at (3178.89, 14547.50). For $k_2=0.15$, it is (4170.96,15998.50).

In Fig.10 we concisely present a comparison between the Gillespie simulation algorithm (SSA) and the SDE simulation in the case of $k_2=0.15$, the comparison being about proportions of large circles in whole dynamic processes. These stemmed from $t-x$ data of the two simulation methods. If a peak is higher than 11855, which is the 85% of the maximum of the unstable manifold on $x$ coordinate, this peak will be regarded as representing a large circle. Parts of a large ring above 1391 are counted as the running time of the large ring, where 1391 is 10% of the maximum value of the unstable manifold. In addition, if a large circle is connected to a limit cycle, the connection will be cut off at 4171, which is the $x$ coordinate of the unstable focus. Though there may be a situation that a circle, whose peak is higher than 13909, does not pass across the barrier to be an intact dynamic process, it will still be considered as a large circle, because it indeed performs the function of the large circle. The parameters, $v=1/3, 1/6, 1/9, 1/12, 1/15$ in Gillespie algorithm, correspond to $D=1,3,5,7,9$ in SDE simulation respectively. What we can see is that with the change of parameters, development trends of proportions of large circles are quite consistent. The capacity of large circles to compete with stable limit cycles in these processes becomes stronger and stronger, being reflected as that proportions of large circles extend with parameters increase.

## Entropy production and heat dissipation

The well-known and widely used definition of entropy in the interdisciplinary of statistical physics and derivative is as follows:

FIG.10. Proportions of large circles in whole dynamic processes of SDE simulation and SSA.

$$S = k_B \int_\Sigma P(x,y,t)\ln P(x,y,t) d\Sigma \qquad (11)$$

where $k_B$ is the Boltzmann constant, and $\Sigma$ is the domain of the system we studied. Thereupon, the time change rate of the entropy $S$ is [39]

$$T\frac{\partial S}{\partial t} = k_B \int_\Sigma (1+\ln P)\nabla \cdot \boldsymbol{J} d\Sigma$$

$$= -\int_\Sigma (k_B T \nabla \ln P - \boldsymbol{F})\cdot \boldsymbol{J} d\Sigma - \int_\Sigma \boldsymbol{F}\cdot \boldsymbol{J} d\Sigma$$

$$= \int_\Sigma \boldsymbol{\Pi}\cdot \boldsymbol{J} d\Sigma - \int_\Sigma \boldsymbol{F}\cdot \boldsymbol{J} d\Sigma$$

$$= EPR - HDR \qquad (12)$$

In eq.12, $EPR$ is the entropy production rate, and $HDR$ is the heat dissipation rate of which the original definition is the time derivative of the expected value of work done by deterministic generalized forces expressed in the form of Stratonovich integral. The introduction of temperature $T$ is to maintain thermodynamic consistency based on Einstein's relation $\boldsymbol{DD}^T=2k_BT$. Therefore, the thermodynamic force $\boldsymbol{\Pi}=\boldsymbol{F}+\boldsymbol{D}\cdot\nabla U$ [39,48]. If the system is in NESS, $\boldsymbol{\Pi}=\boldsymbol{J}_{ss}/P_{ss}$.

In NESS, we have

$$T\frac{\partial S}{\partial t} = \int_\Sigma (\boldsymbol{J}_{ss}/P_{ss} - \boldsymbol{F})\cdot \boldsymbol{J}_{ss} d\Sigma = \int_\Sigma \boldsymbol{D}\nabla U \cdot \boldsymbol{J}_{ss} d\Sigma = 0 \qquad (13)$$

The most direct inference is that $\boldsymbol{J}_{ss}$ is orthogonal to $\nabla U$. This can only be proved when $\boldsymbol{D}\to 0$ [49]. It can be explained vividly that the direction of $\boldsymbol{J}$ can be determined by the deterministic generalized force $\int_\Sigma \boldsymbol{F}\cdot \boldsymbol{J} d\Sigma dt$ only when the noise almost does not exist. The general explanation is that $\nabla\cdot\boldsymbol{J}=0$, $\boldsymbol{J}$ is the curl flux, and the integral along any closed curve (i.e. the work done by $\nabla U$) in the gradient field is zero.

Since the entropy production rate $EPR$ is equal to the heat dissipation rate $HDR$ in NESS, we only need to calculate the heat dissipation rate which is easy to calculate, then we can get all the information of the two. Fig.11 shows the heat dissipation rates of $k_2=0.12$ and $k_2=0.15$.

For $k_2=0.12$, the polyline of $HDR$ is similar to that in other researches, which shows certain monotonicity and increases with the increase of the diffusion coefficient $D$ [26,42]. Additionally, for $k_2=0.17$, the shape of the polyline is virtually identical to the result of $k_2=0.12$. For $k_2=0.15$, the polyline presents the shape of a checkmark, which is not common that there is a minimum value of $HDR$ in the neighborhood of $D=3$. What is the essence of entropy production rate and heat dissipation rate? What does the minimal value of the heat dissipation rate imply (Fig.11(b))? An intuitive illustration, which is that the increase of noise makes a part of $J_{ss}$ shift from stable limit cycle to large circle, can be seen in *SI* text. However, we prefer to clarify its physical significance in the following.

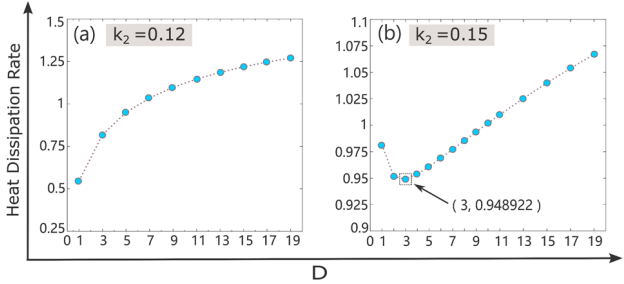

FIG.11. Heat dissipation rates of $k_2=0.12$ and $k_2=0.15$. In (b), $k_2=0.15$, we reduce the interval between the points of diffusion coefficients $D$ from 1 to 10, the reason of which is to make the minimal value of HDR more evident. In our calculation, $P(x,y,t)$ of each $D$ has been normalized.

According to eq.(8), we define $J=v_P P(x,y,t)$, in which $v_P=F'$ as the velocity of probability flow. Meanwhile, $F$ from the overdamped Langevin equation is still interpreted as a generalized force. In the phase space dynamics determined by F-P equation and described by probability, the velocity of the state should be represented by $v_P$. In this expression, $dx/dt=v_P$. Therefore, the heat dissipation rate is

$$HDR=\int_\Sigma F\cdot v_P P d\Sigma \qquad (14)$$

Eq.(14) has the form of power. It connotes that we may explicate $HDR$ from the perspective of work. However, the self-consistency of this idea initiated by $v_p$ needs to be demonstrated before we continue to analyze it. One can see the proof in *SI* text.

The material derivative does not comprise the change of entropy produced by the motion of the state point (eq.16), since in the whole process from the initial state to the non-equilibrium steady state, the deterministic generalized force $F$ only plays the role of maintaining the movement of the state point. Its work, which is, is immediately dissipated by the overdamped system. From a global perspective, as a deterministic and constant force field, $F$ pushes the dynamic point motion according to its established structure from beginning to end. $F$ really consumes the energy supply of the system (in biological systems, the energy supply is chemical energy), which is reflected in the kinetic behavior of the system in the non-equilibrium state. This is the same whether it is in the NESS or before reaching the NESS.

On the contrary, the stable dynamic behavior of the system does not need to be continuously driving by the stochastic action $D\nabla U$, because the expected value of the diffusion motion driven by white noise is zero. Before reaching NESS, the result of its work is to make the state point reach the position where the deterministic force cannot realize, which is the process of constructing the landscape (the dynamic structure of phase space) from the initial state. After reaching NESS, the development of landscape has been completed, and the weighted average of $D\nabla U\cdot v_p$ is zero, which means that the stochastic forces cancel each other, which is the character of thermal fluctuation.

In terms of probability, the deterministic force field is constant, which does not provide new possibilities for the state of the system. When the system reaches NESS, the landscape does not change, so the stochastic effect is zero. This is consistent with the explanation of entropy in statistical physics.

Through the above analysis, we know that heat dissipation is a measure of system energy consumption, which has the meaning of "power". This conclusion applies to all systems which can be described by Eq.3 and Eq.4. For a biochemical system, such as a set of genetic circuits, the energetic material is generally ATP, and the stochastic effect comes from the hot bath, the randomness of biochemical reaction, or the copy noise of plasmids, and so on. For ecosystems, the energy supply materials may be producers and primary consumers, and the sources of noise are more extensive. No matter what kind of system, the explanation of Fig.11(b) is the same: the minimum value of heat dissipation rate means that the system has a relatively minimum energy consumption rate, that is, the energy consumption required to maintain the system in NESS is the least.

## Application to a practical model

Now we discuss a practical biological model. The model comes from [50]. Its main content can be described by equations similar to eq.(1) and (2):

$$\frac{dx}{dt}=v+\frac{0.01gx^2}{r^2+0.01x^2}-\frac{0.1x}{1+0.1x+y}$$

$$\frac{dy}{dt}=\frac{h}{1+10^{-5}k^5x^5}-\frac{y}{1+0.1x+y} \qquad (22)$$

where the parameters of this set of equations come from the supplemental material of [50]. $k=4.5045$ is the coefficient of regulating Hill function, which is invariant normally. By adjusting parameters $g$, $h$, $r$ and $v$, one can obtain different dynamics structures of this model. We chose two sets of parameters to study. One set is $g=0.043$, $h=0.974$, $r=0.3$, $v=0.00028$. The system controlled by these parameters will be similar to the case in our work when $k_2=0.12$. That is, in the first quadrant, there are three fixed points in the system, which are stable fixed point, saddle, and unstable focus (Fig.12(a)). Another set is $g=0.106$, $h=0.978$, $r=0.4$, $v=0.00016$. For equations controlled by this set of parameters, the system will present similar to the case in our work of $k_2 = 0.17$. Three fixed points were found in the first quadrant, which is the stable fixed point, saddle, and stable focus (Fig.12(b)).

With the noise being introduced in this system, we can see in Fig.12(c) that a large circle along with the unstable manifold is formed. It means that this actual biological system can also appear noise induced quasi-periodicity, and even switch between different periods by reasonably adjusting the parameters, i.e. reaction rates. To some degree, this shows that the quasi-period induced by noise is common, and the results of our work are of general significance.

Naturally, the same analytical techniques and theories in our work can also be used in this model. Application. In addition, in *SI* text, we explore the mechanism of this seemingly universal structure.

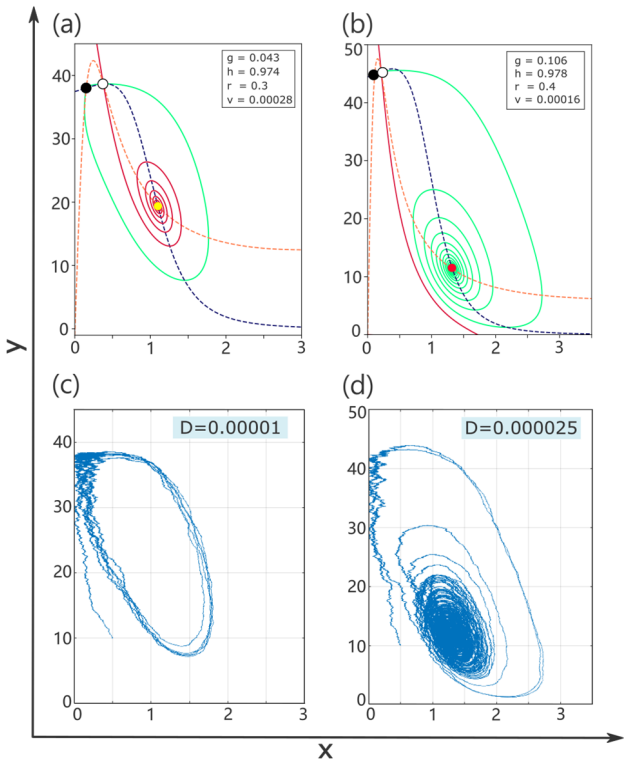

FIG.12. Dynamics structures and stochastic streamlines of the model coming from [50]. The use of the color and line type is the same as Fig.3. In (a) and (b), the black solid circle represents the stable fixed point and the black hollow circle for the saddle. In (a), the yellow solid circle represents the unstable focus. In (b), the red solid circle represents the stable focus. (c) and (d) are stochastic streamline diagrams of the cases described in (a) and (b) respectively.

## Discussion

In this work, we utilized a classical model with positive and negative feedbacks, producing three kinds of dynamic structures by changing a key parameter $k_2$. We thoroughly studied two structures of them, one of which is the *fixed point-saddle-unstable focus* structure produced a quasi-period by noise ($k_2 = 0.12$), another of which is *fixed point-limit cycle* structure produced the switch between two quasi-periods by noise ($k_2 = 0.15$).

Now we can try to answer the questions raised at the beginning of this article. Noise can induce the system to produce a new periodic state, which is different from the transition between bistable states or the relatively long stay near the unstable fixed point (similar to the case of $k_2 = 0.17$), and which is independent of the critical phenomenon near the bifurcation. In the case $k_2 = 0.12$, noise induces a new quasi-periodic behavior resembling a *saddle node on an invariant cycle* (SNIC) bifurcation, which does not exist in the original system [51]. For $k_2 = 0.15$, the new periodic behavior induced by noise coexists with the limit cycle in the original system and competes with each other, resembling a *saddle homoclinic orbit* (HOM) bifurcation based on an intrinsic stable limit cycle having existed [51]. This can only happen in a stochastic process.

As long as the deterministic dynamic structure of a system has the characteristics of the model we studied, the periodicity induced by noise has the conditions to realize, just as the biological example given in the previous paper. In fact, similar genetic circuits can be synthesized artificially in synthetic biology [52-55]. By properly adjusting the rates of biochemical reactions, that is, the parameters in the differential equations, a structure that meets our requirements can be obtained.

For the case of $k_2 = 0.12$, the periodicity based on the large circle can be selectively activated by adjusting the noise intensity. The sensitivity of the system to noise can be realized by adjusting the distance between the saddle point and the node. If our research object is a biochemical system, such as the genetic circuit, this adjustment is to change some reaction rate. For $k_2 = 0.15$, it is worth noting not only the noise-induced large circle, but also the switch and competition between the two cycles. Especially, when the noise increases, the motion on the stable limit cycle becomes disordered, and the frequency of the large cycle increases. This may mean that in some systems, the large circle is an auxiliary mechanism to partly replace the unstable original periodic structure.

Period switch and competition are also reflected in nature. The periodical cicadas are a good instance [56,57]. There are many explanations about the life cycle of this insect, such as the seasonal cycle of the host [58], resource constraints [59], the influence of the ice age paleoclimate [60], and so on. Other work shows different views and mentions the mutation of the cycle [61]. Various reasons lead to periodical cicada forming a set of accurate time counting mechanism, and external factors such as climate, resources, and predators are the external factors that affect the timing mechanism, such as environmental noise. We might as well speculate that it may be a genetic circuit with unclear structure, which controls the life cycle of the periodical cicadas.

In fact, the life cycle of 17-year cicadas is longer than that of 13-year cicadas by four years inhibition period of early pupal growth. Reference [62] proved the homology between 13-year cicadas and 17-year cicadas by measuring the mitochondrial DNA of periodical cicadas, and this work negated the hybridization hypothesis, concluding that it is only because 17-year cicadas experienced a rapid development period of 4 years in the development process. In terms of dynamics, if the extra four years are regarded as the time taken by the state point to cross a potential barrier, then the model similar to our work can be a candidate. In addition to the model of $k_2 = 0.15$ which can be used to explain the dynamics, it is also possible to use the $k_2 = 0.12$ model for an explanation, considering that the starting points of the cycle of 17-year cicadas and 13-year cicadas are respectively before and after the barrier. Moreover, it is also an available explication that the existence of the barrier or not respectively represent the 17-or-13 years cycle, in which the barrier can be produced or removed by regulating the coefficient $a$ (see *SI text* for details).

In the model of this work, the constant $a$ in $F_1(x, y)$ represents the ratio of the formation rate of the molecule $x$ to the degradation rate in the equation (eq.1). In other words, the process of $\alpha$ decreasing can be seen as the process that the ratio of 'birth rate' to 'death rate' of $x$ getting smaller and smaller. In this way, as the ratio decreases, the original stable periodic structure diverges. The excessive attenuation of $x$ caused by noise will result in the number of both $x$ and $y$ dropping to near-zero rapidly. Then, a new cycle will be started stochastically, as described in Fig.S5(c,d), Fig.S6(c,d), and Fig.S7(c,d). We expect that this will help to explain the dynamics of some species evolution.

We expect that the results of this work can be verified in experiments and real world. Further work should focus on whether it can set an effective switch to control the switching between two periods, or even turn on one period and turn off another. In addition, the significance of the minimum value of heat dissipation rate in the actual system is also worth exploring in the actual system.

# Appendix

The parameters in eq.(1) and (2) were initially derived by a set of biochemical reactions, being the reaction rates. They may have been determined by experiment at that time, plus some artificial assumptions. No matter what, it hardly affects the result. In this work, parameters are:

Eq.(1): $\alpha_1 = 0.00875$, $\beta_1 = 7.5$, $k_1 = 2.5 \times 10^7$, $\delta = 4 \times 10^{-8}$, $\lambda_1 = 0.0004$, Hill coefficient $n = 2$;

Eq.(2): $\alpha_2 = 0.025$, $\beta_2 = 2.5$, $\lambda_2 = 0.0004$, Hill coefficient $p = 5$, $k_2$ is the key parameter that changes according to the requirements.

These parameters are dimensionless in our work, which also reflects our original intention: we hope this work can benefit not only genetic circuit